\documentclass[12pt]{article}
\pdfoutput=1
\usepackage{graphicx}
\usepackage{subfigure}

\newcommand\pubnumber{SNSN-323-63}
\newcommand\pubdate{\today}

\def\hok{1.Hokkaido University\\
5tyoume, Kita8-jonishi, Kita-ku Sapporo-shi, Hokkaido, 060-0808, Japan}
\def\kek{2.High energy accelerator research organization, KEK\\
1-1, Oho, Tsukuba-shi, Ibaraki, 305-0801, Japan}
\def\osaka{3.Osaka University\\
1-1, Yamadaoka, Suita-shi, Osaka, 565-0871, Japan}

\def\Title#1{\begin{center} {\Large #1 } \end{center}}
\def\Author#1{\begin{center}{ \sc #1} \end{center}}
\def\Address#1{\begin{center}{ \it #1} \end{center}}

\newcommand\pubblock{\rightline{\begin{tabular}{l} \pubnumber\\
         \pubdate  \end{tabular}}}
\newenvironment{Abstract}{\begin{quotation}  }{\end{quotation}}
\newenvironment{Presented}{\begin{quotation} \begin{center} 
             PRESENTED AT\end{center}\bigskip 
      \begin{center}\begin{large}}{\end{large}\end{center} \end{quotation}}
\def\Acknowledgements{\bigskip  \bigskip \begin{center} \begin{large}
             \bf ACKNOWLEDGEMENTS \end{large}\end{center}}

\def\beq{\begin{equation}}
\def\eeq#1{\label{#1}\end{equation}}
\def\eeqn{\end{equation}}

\def\beqa{\begin{eqnarray}}
\def\eeqa#1{\label{#1}\end{eqnarray}}
\def\eeqan{\end{eqnarray}}

\let\bar=\overbar

\def\Dslash{\not{\hbox{\kern-4pt $D$}}}
\def\dslash{\not{\hbox{\kern-2pt $\del$}}}

\def\msb{{\bar{\ssstyle M \kern -1pt S}}}

\begin{document}
\begin{titlepage}
\pubblock

\vfill
\Title{SOI pixel circuits with synchronized TMC for
time-of-flight stigmatic imaging mass spectrometry}
\vfill
\Author{ Kaori Watanabe$^1$, Masayuki Ikebe$^1$,Youichi Fujita$^2$, Yasuo Arai$^2$, and Hisanao Hazama$^3$}
\Address{\hok}
\Address{\kek}
\Address{\osaka}
\vfill
\begin{Abstract}
We propose SOI pixel circuits with a synchronized time memory cell
(TMC) for time-of-flight stigmatic imaging mass spectrometry. The
circuits simultaneously detect the position and the fine/coarse flight
time of an ion for the MALDI-ToF mass spectrometer. We discuss the
circuit design and present the simulation results of a prototype
detector comprised of a 32$\times$32 pixel array in which each pixel pitch
is 40 $\mu$m  and the time resolution is a minimum of 1 ns. The results of
transient analysis demonstrate the fully correct synchronous operation
at a 100-MHz clock frequency and simultaneous 32-word SRAM writing.
\end{Abstract}
\vfill
\begin{Presented}
International Workshop on SOI Pixel Detector\\
Sendai, Japan, June 3--6, 2015
\end{Presented}
\vfill
\end{titlepage}
\def\thefootnote{\fnsymbol{footnote}}
\setcounter{footnote}{0}

\section{Introduction}
Imaging mass spectrometry (IMS), which is combination of mass separation and imaging, has emerged as a powerful tool in fields such as pathology, pharmacology, and others. Matrix-assisted laser desorption/ionization (MALDI) and time-of-flight mass spectrometry (TOFMS) are the main forms of IMS. The stigmatic method of MALDI-IMS has attracted a particularly great deal of attention because spatial resolution is not due to laser spot size and the measurement time can be reduced by the 2-D ion detector\cite{hazama_development_2011,aoki_novel_2011}. In this method, MALDI produces a 2-D projection with multiple ions, where an ion optical lens expands the ion image and the 2-D ion detector obtains the positions and time-of-flight (ToF) of the ions. The stigmatic imaging mass spectrometer thus requires a multi-hit detector to create an ion ToF imaging. In this work, we propose SOI pixel circuits with a synchronized time memory cell (TMC) for time-of-flight stigmatic imaging mass spectrometry. The circuits simultaneously detect the position and the fine/coarse flight time of an ion for the MALDI-ToF mass spectrometer. We discuss the circuit design and present the simulation results of a prototype detector comprised of a 32 $\times$ 32 pixel array in which each pixel pitch is 40 $\mu$m and the time resolution is a minimum of 1 ns.

\begin{figure}[htb]
\centering
\includegraphics[height=2.0in]{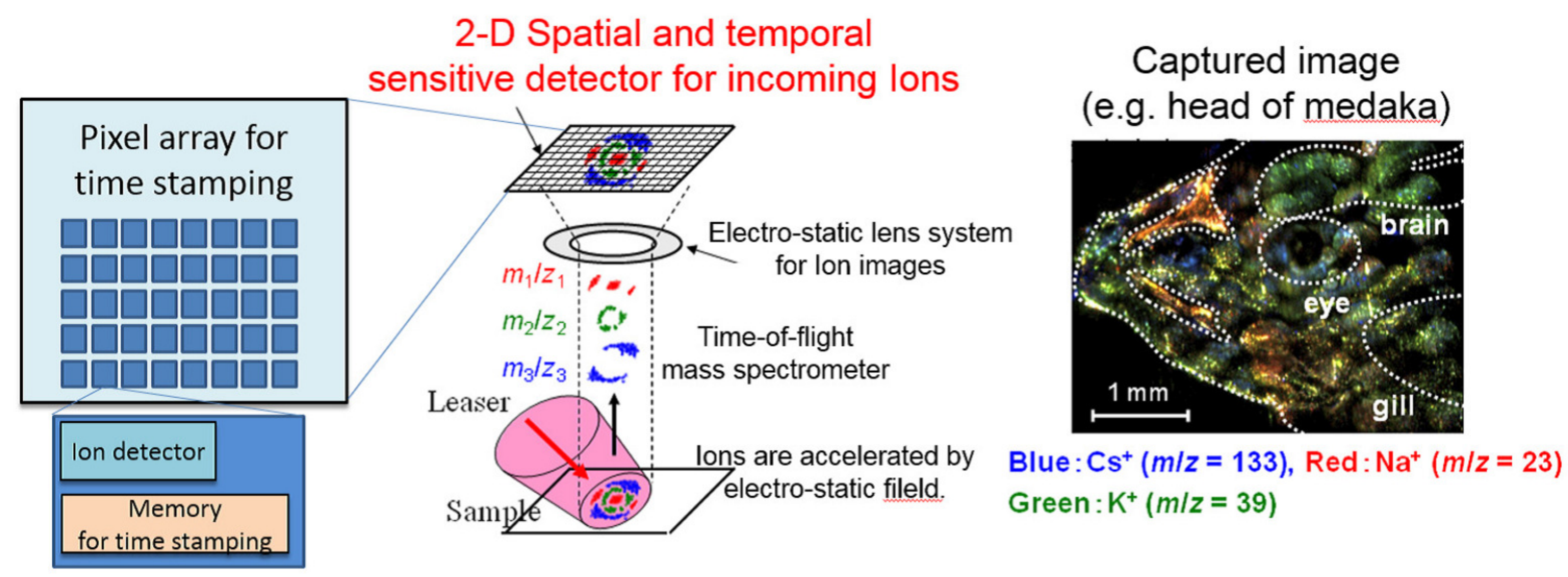}
\caption{Time-of-flight stigmatic imaging mass spectrometry.}
\label{fig:magnet}
\end{figure}

\section{Concept of our SOI detecter}

The concept of our SOI detector is shown in Fig. 2. The sensor consists of a pixel block, column parallel A/D converters, counters, and clk/reference signal generators. The pixel circuits are composed of ion detectors, flag generators, analog S/H circuits, D-FFs, and latches for TMC operations. Clk and reference signal generators include a quadrature oscillator and a multi-phase clock generator that generate 8-MHz $sin$/$cos$ waves and 4-phase 2-MHz clock signals, respectively. The analog S/H circuits store the states of $sin$/$cos$ waves with 6-bit resolution (= 1-ns resolution), the TMCs store the states of the 4-phase signals with 3-bit resolution (= 62.5-ns resolution), and the D-FFs store the counter values with 5-bit resolution (= 500-ns resolution). The whole resolution of ToF is 12 bits and the LSB has a resolution of less than 1 ns. All pixels store the 12-bit ToF data using a common counter and signal generator.

\begin{figure}[htb]
\centering
\includegraphics[height=2.0in]{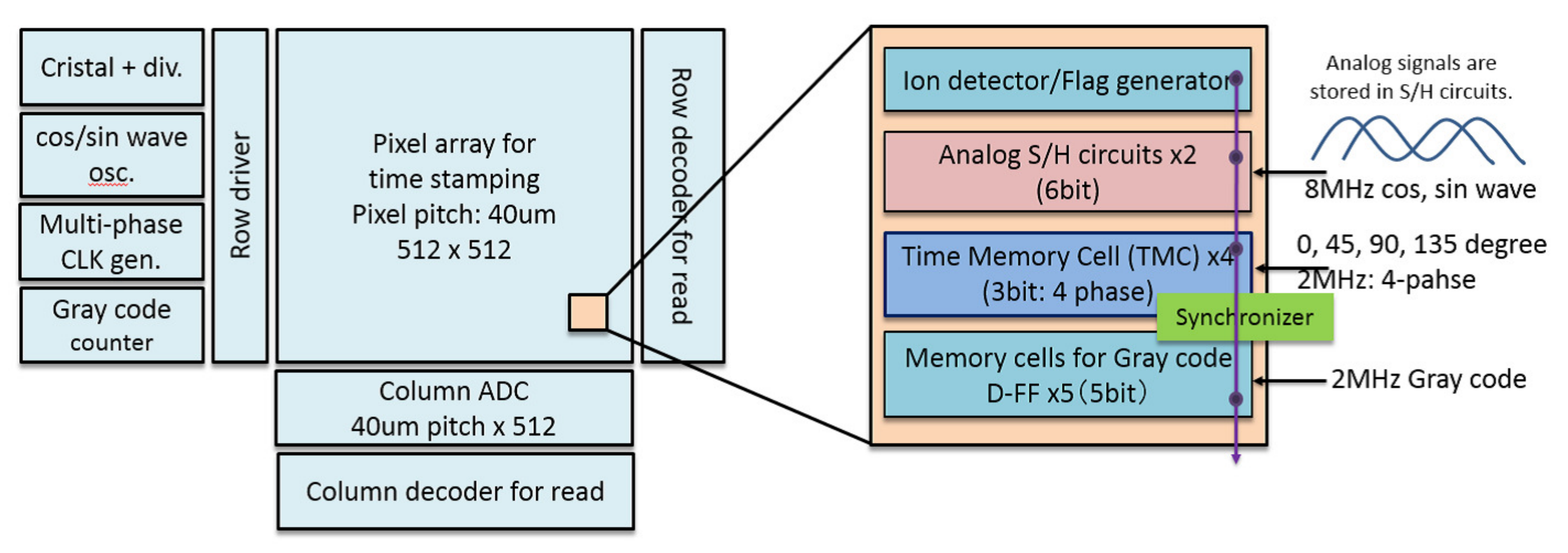}
\caption{Concept of SOI detector.}
\label{fig:magnet}
\end{figure}

\begin{table}[h]
\begin{center}
\begin{tabular}{lc}  
\multicolumn{2}{c}{Specifications}     \\ \hline
 Pixel pitch (res.)  &   40 $\mu$m (32 x 32)   \\
Time resolution  &   1 ns (6 bit ADC)   \\
 TMC resolution  &   62.5 ns (3 bit)   \\
 Counter resolution  &   500 ns (5 bit)   \\
 CLK. Freqency  &   2.0 MHz   \\
\hline
\end{tabular}
\caption{SOI detector specifications.}
\label{tab:blood}
\end{center}
\end{table}

\section{SOI pixel circuits with synchronized time memory cell (TMC)}
\subsection{Time memory cell}
A time memory cell (TMC) was proposed by Y. Arai et al. and applied to measure the drift times of electrons\cite{arai_time_1998}. 
This cell harnesses the low-power and high-density characteristics of a CMOS memory cell and the short delay time of a gate. The TMC also functions as a time to digital converter (TDC) and stores an input-signal-falling/rising timing by using fixed delay signals generated from a PLL. Usually, the TMC uses the fixed delay signals to store the timing of input falling/rising edges. However, in our circuits, we use the fixed delay signals, which are multi-phase clocks, as the input data and store them by the input falling/rising edges. Since one clock period is separated into edges of multi-phase clock signals, the TMC achieves fine time measurement within one clock cycle. For example, the n-bit counter-based time to digital convertor (TDC) requires 2$^{n}$ clock cycles, i.e., combined usage of the m-bit TMC, to reduce conversion time to 2$^{n-m}$ clock cycles. In this study, we use the TMC for fine time to digital conversion and to relax the clock rate. Relaxing the rate leads to stable operation of the large detector.

\begin{figure}[htb]
\centering
\includegraphics[height=2.0in]{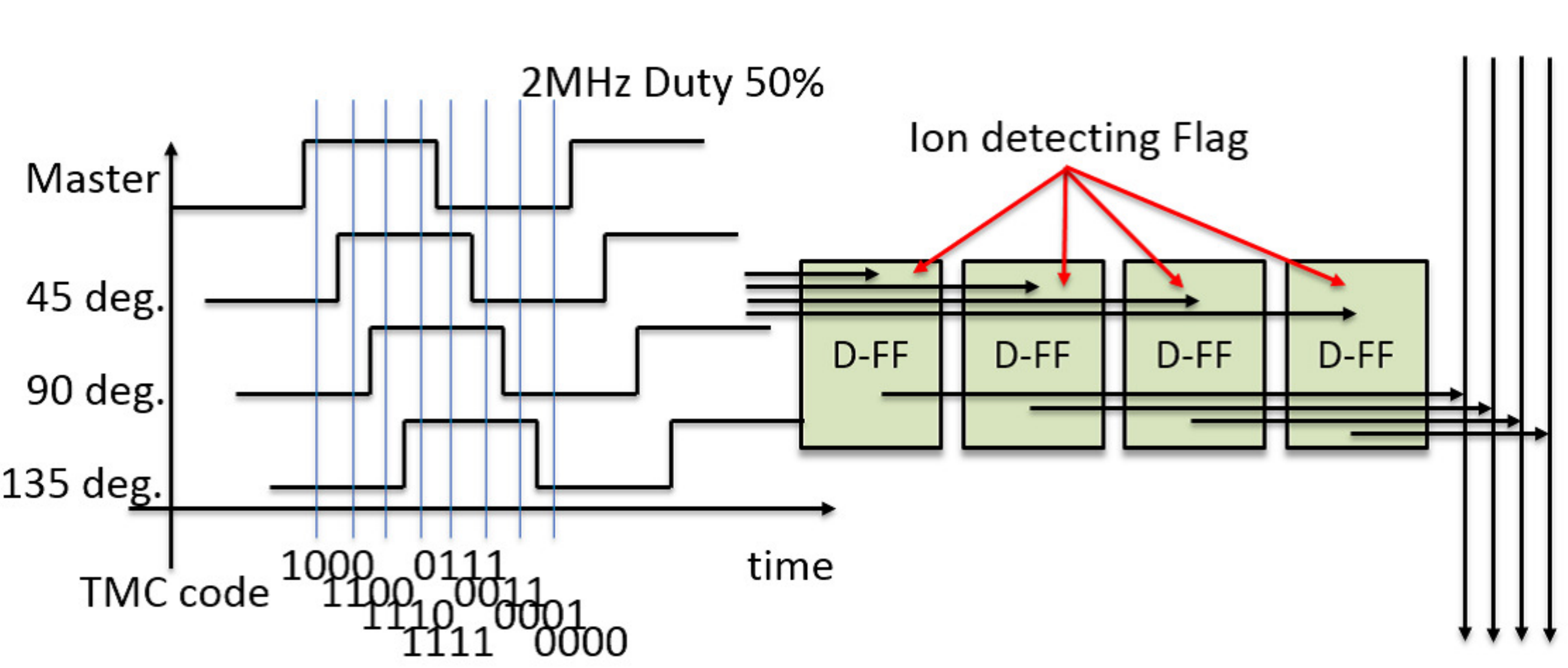}
\caption{Time memory cell (TMC) with multi-phase clock.}
\label{fig:magnet}
\end{figure}

\subsection{Synchronization and consistency issues with TMC}
In the combined usage of the TMC and the counter-based TDC, special care should be taken with the causality of the TMC and the counter\cite{takahashi_digital_2011,uchida_12-bit_2014}. By `causality' we mean that the lower bits (TMC output) have to be coupled with the upper bits (counter output). In other words, the settled TMC state completely decides the counter state with synchronization. Conventionally, the shared clock is given to a TMC and a counter for the causality. However, when the falling edge of the stop signal (pixel flag of ion detection) is close to the clock signal timing, metastability occurs and these circuits work at individual timings. So, even using a shared clock, the causality degrades due to the metastability. Delay adjust cannot fundamentally handle the metastability, either. Figure 6 shows an example of measured miscoding probability while the stop signal is close to the clock signal. We confirmed a miscode generation of 50$\%$ when the stop-signal timing was equal to the master clock. Exact causality is important for robustness against metastability.

\begin{figure}[htbp]
\begin{tabular}{cc}
\begin{minipage}{0.5\hsize}
\centering
\includegraphics[height=1.2in]{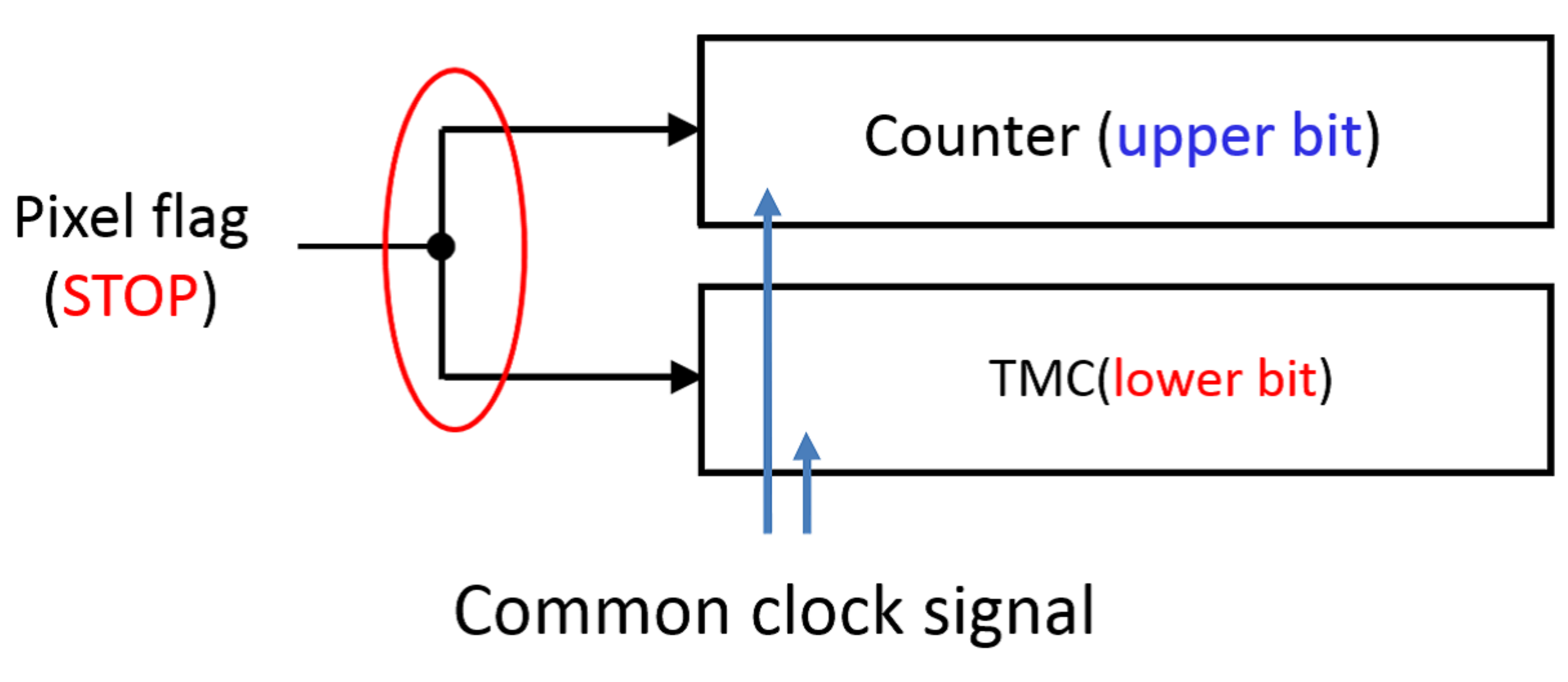}
\caption{Conventional TMC with counter architecture.}
\label{fig:magnet}
\end{minipage}
\begin{minipage}{0.5\hsize}
\centering
\includegraphics[height=1.2in]{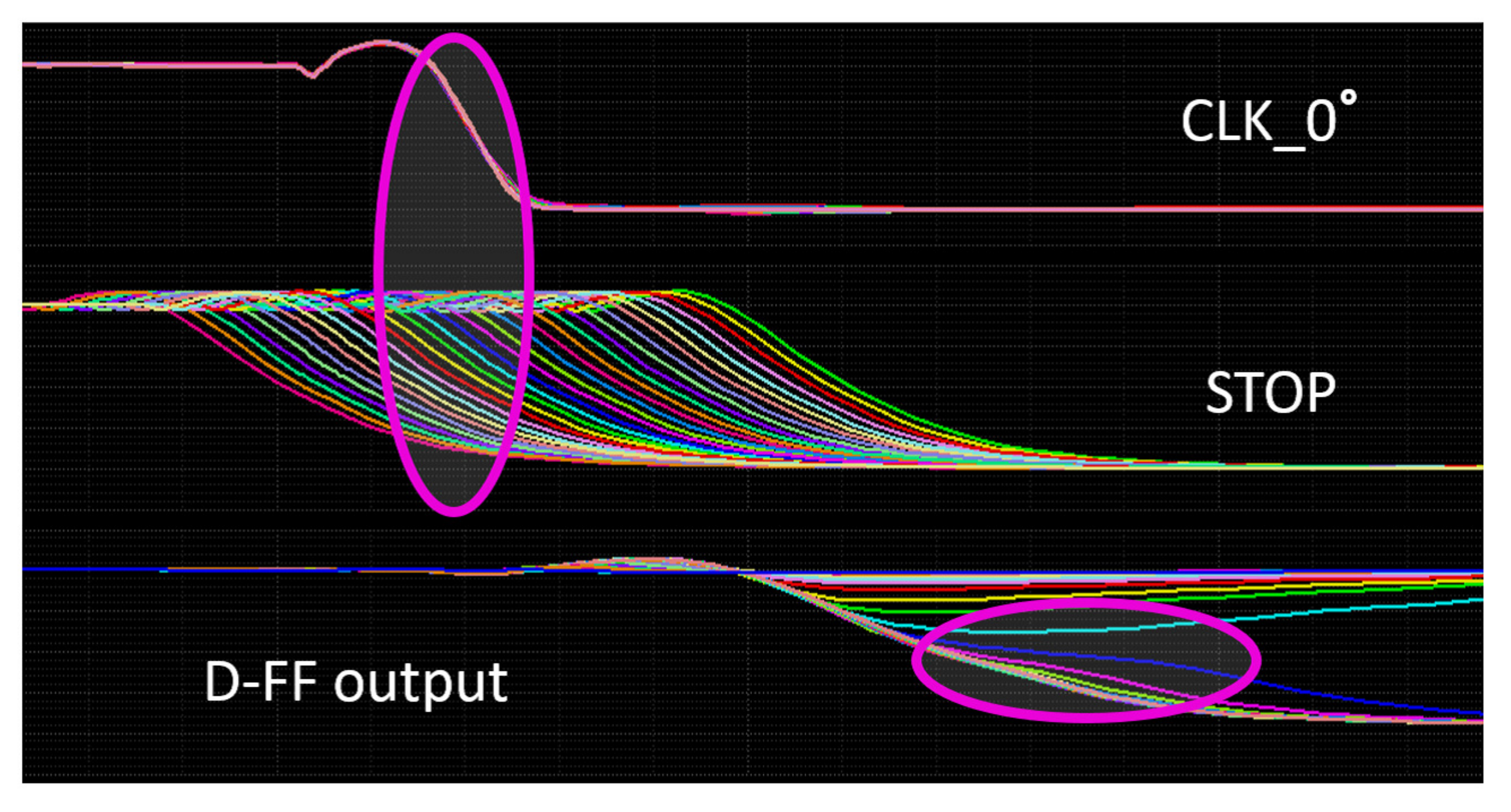}
\caption{Simulation results of metastability analysis.}
\label{fig:magnet}
\end{minipage}
\end{tabular}
\end{figure}
\begin{figure}[htb]
\centering
\includegraphics[height=1.5in]{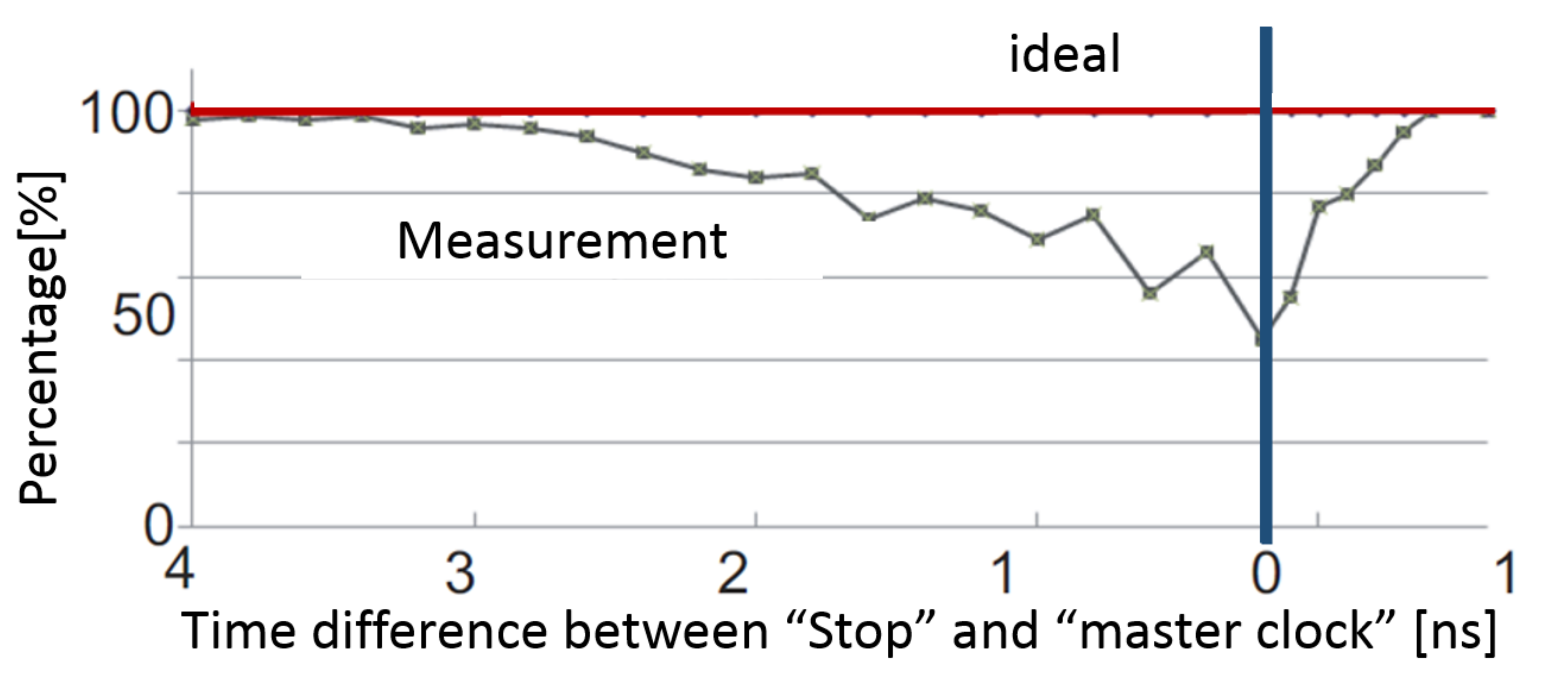}
\caption{Miscoding probability $(\%)$.}
\label{fig:magnet}
\end{figure}

\section{Proposed circuits}
\subsection{Proposed synchronized TMC}
Figure 7 shows the block diagram of the proposed synchronized TMC. Our synchronized TMC with a counter consists of memories, a Schmitt trigger, and a common Gray code counter. For 3-bit resolution of the TMC, 4-phase clock signals are given to each D-latch. Only a master clock without phase shift is used for storing counted values through the first D-FF of the TMC (TMC$\langle$0$\rangle$). In this configuration, the settled state of the TMC$\langle$0$\rangle$ by the stop signal completely holds the counter state. Figure 8 shows internal states of the D-FFs for the counter output. When the settled state of the TMC$\langle$0$\rangle$ is ``0'' and the master clock is also ``0'', the counter keeps its previous state $T_{n-1}$. Then, when the settled state of the TMC$\langle$0$\rangle$ becomes ``1'' and the master clock is ``1'', the current counter state ``$T_{n}$'' is stored in the D-FFs. Causality between the TMC and the counter can be achieved by using this scheme. However, when the stop signal is close to the clock signal, metastability occurs on the TMC$\langle$0$\rangle$ and miscodes may be generated. We therefore applied the Schmitt trigger between the TMC and the counter to remove the metastability of the TMC$\langle$0$\rangle$ and obtain exact causality/synchronization between the TMC and the counter.

\begin{figure}[htb]
\centering
\includegraphics[height=2.0in]{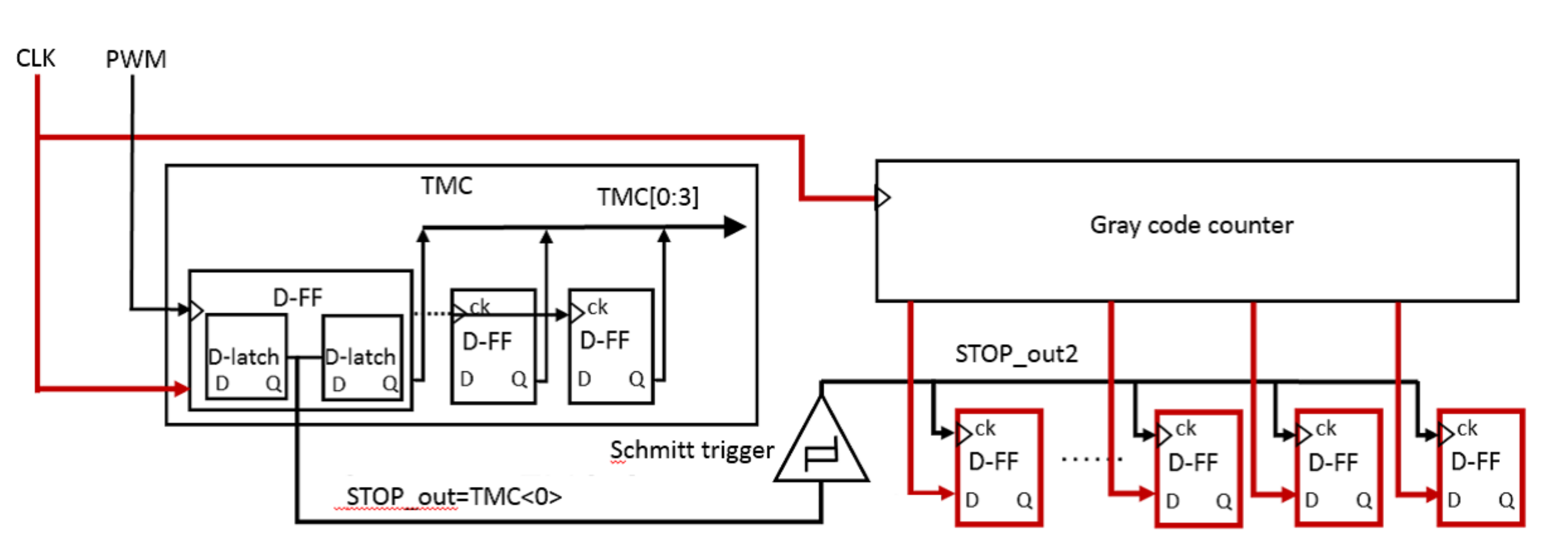}
\caption{Proposed synchronized TMC with counter.}
\label{fig:magnet}
\end{figure}
\begin{figure}[htb]
\centering
\includegraphics[height=1.7in]{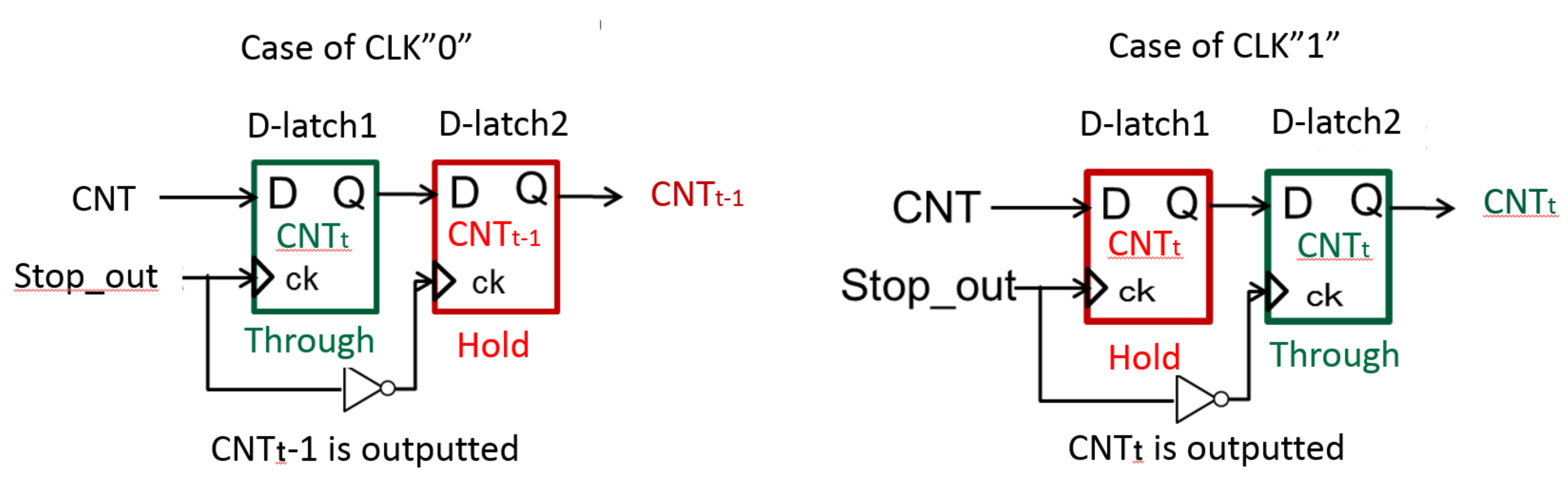}
\caption{Internal state of D-FF of counter output.}
\label{fig:magnet}
\end{figure}

Figure 9 shows simulated delay time by the metastability. Using the Schmitt trigger, we were able to prevent the delay with the signal falling and rising. When the stop signal most closely approached the clock signal, the maximum delay due to metastability was 1.118 nsec. The minimum delay was 0.320 nsec, which was the intrinsic delay of the D-latch. Simulated maximum delay of the counter was 0.959 nsec. The delay of the TMC$\langle$0$\rangle$ D1 should be longer than the counter delay D2 in order to prevent the TMC$\langle$0$\rangle$ from transitioning earlier than the counter-state changing, so we added an extra 1-ns delay to the D1 to satisfy the D1 $>$ D2 condition.

Figure 10 shows our transient analysis of the proposed circuit with metastability. Fully correct operation at the 100-MHz clock frequency when the stop signal was close to the clock signal with 100 fs steps was confirmed.

\begin{figure}[htb]
\centering
\includegraphics[height=2.2in]{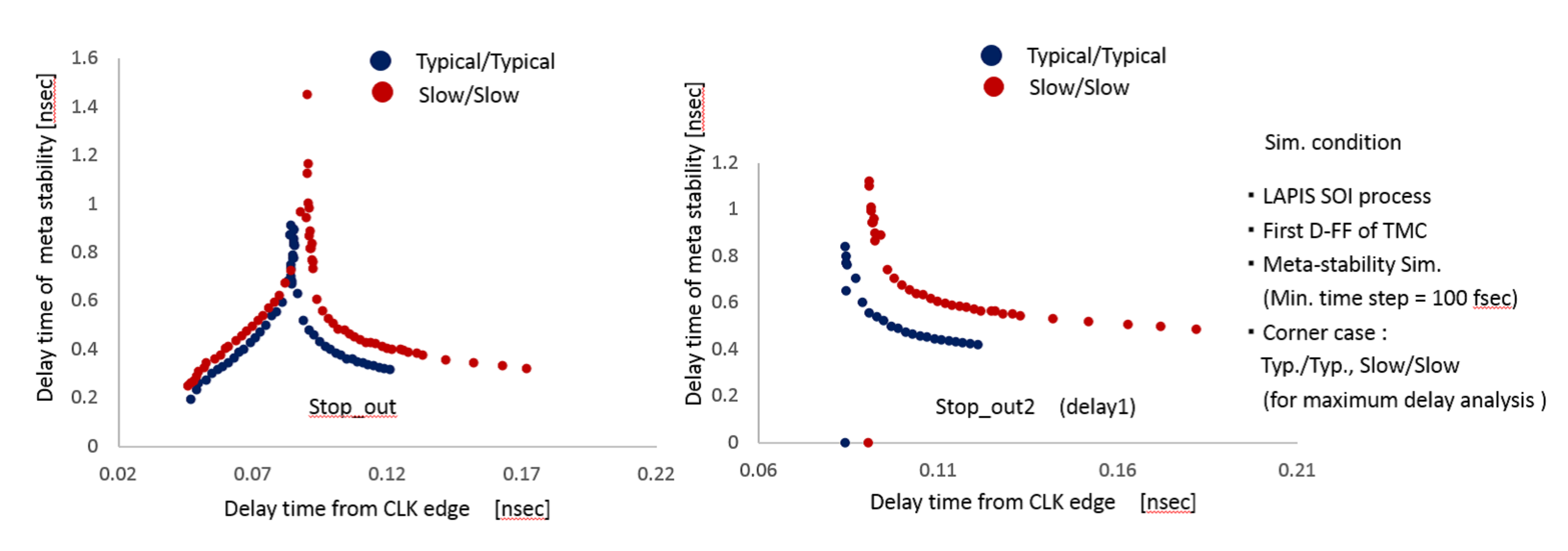}
\caption{ Delay time of metastability.}
\label{fig:magnet}
\end{figure}
\begin{figure}[htb]
\centering
\includegraphics[height=2.2in]{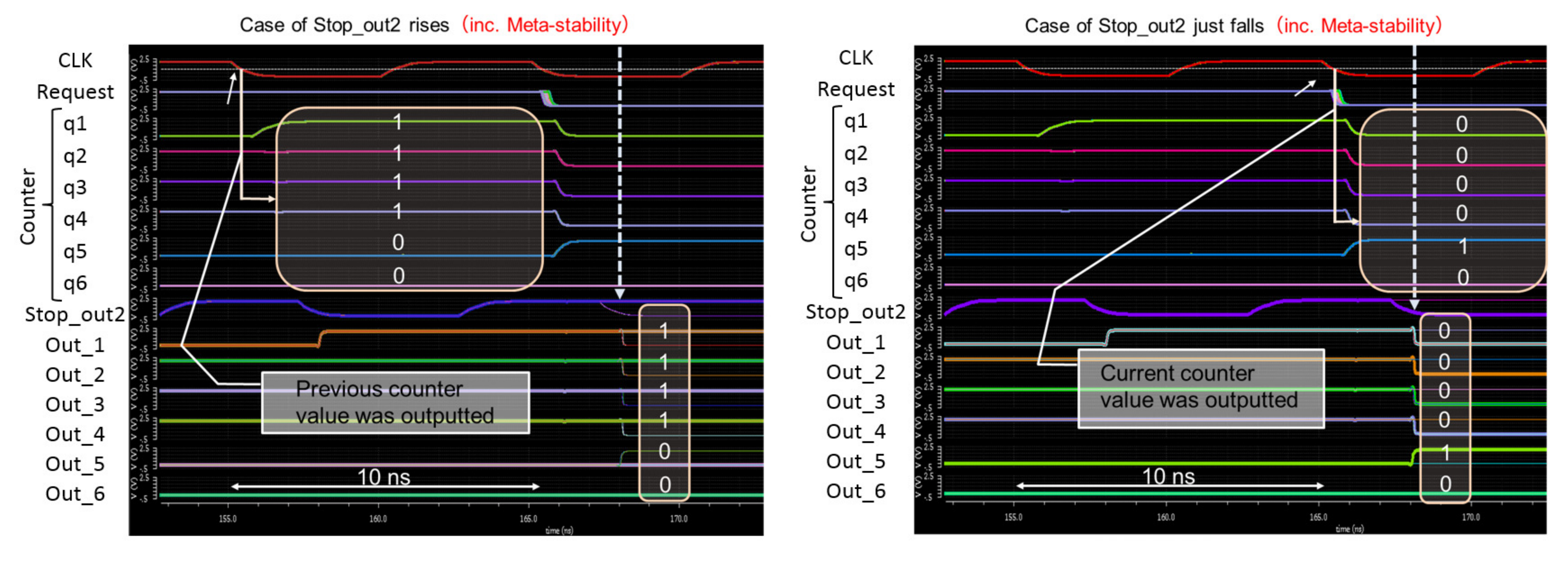}
\caption{Simulation results of proposed circuits with metastability.}
\label{fig:magnet}
\end{figure}
\newpage

\subsection{SRAM cells with dynamic operation}
Being able to handle full multi-hit scenarios is important for ToF stigmatic imaging mass spectrometry. In our design, we used SRAM cells for the memories of the small TMC in the pixel circuits. However, the SRAM cells do not correspond to synchronous multi-writing by their own drive power to bit lines, so we cut the drive power applying dynamic operation to the SRAM cell for the full multi-writing, which is the concern of full multi-hit scenarios. Figure 11 shows our redesigned SRAM cells with the dynamic operation: (a) is a SRAM-based cell and (b), (c) are the dynamic D-latch and D-FF cells, respectively. Each circuit has a quasi two-port structure for independent writing and reading schemes.



\begin{figure}[h]
\centering

\centering
\subfigure[]{\includegraphics[width=3cm]{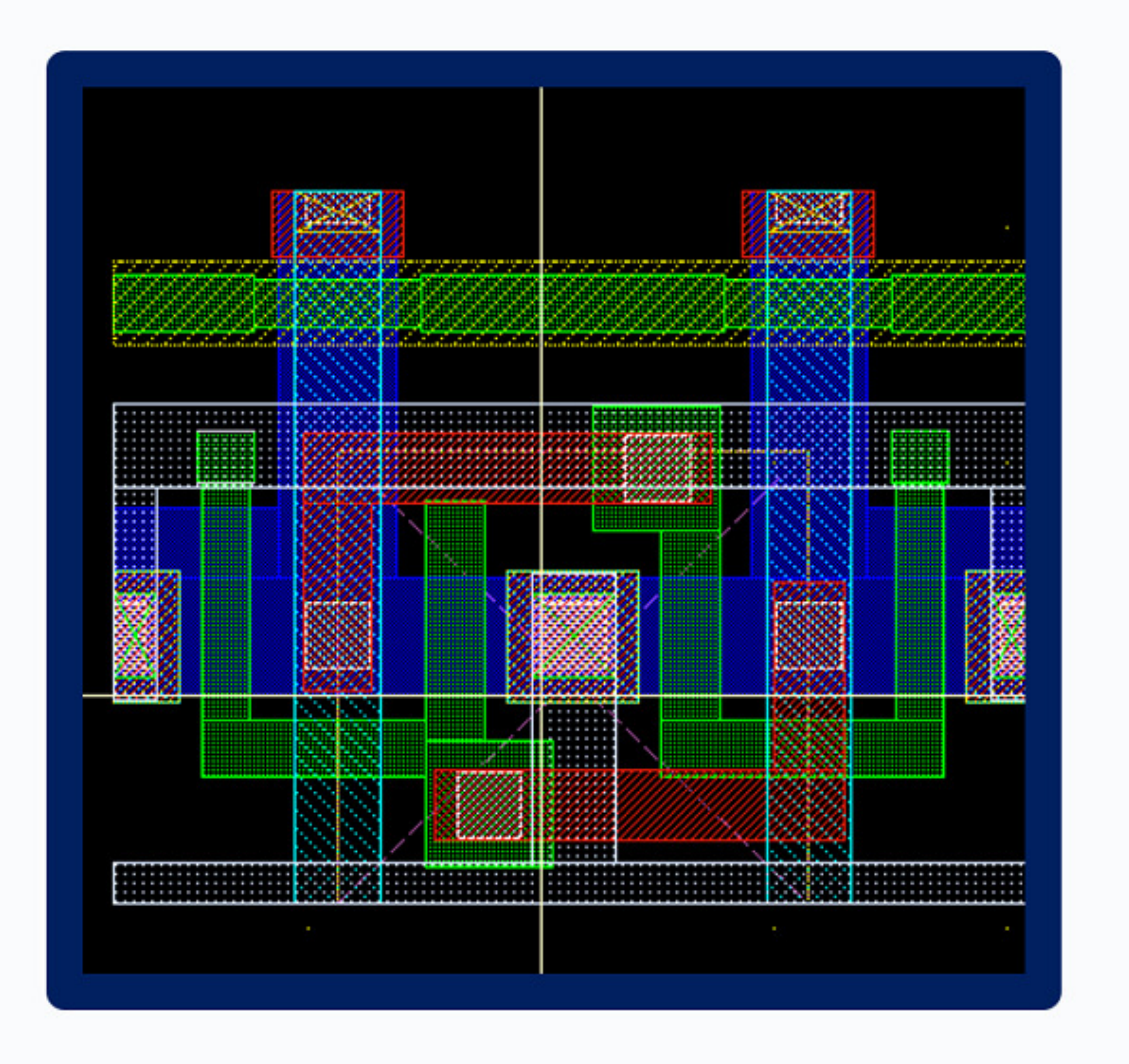}
 \label{fig:fig01left}}
 \hspace{20mm}
\subfigure[]{\includegraphics[width=3cm]{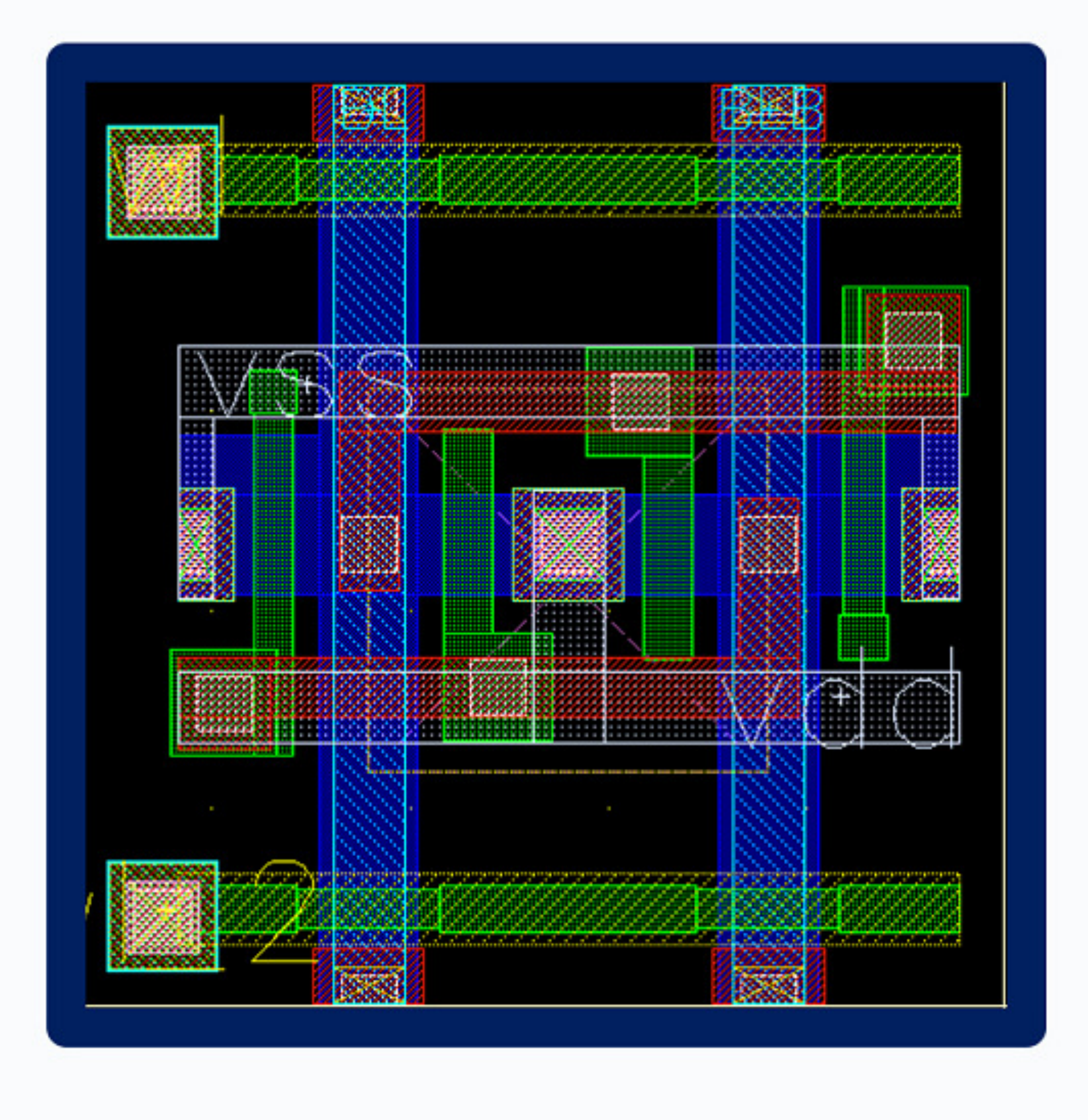}
 \label{fig:fig01left}}
\hspace{20mm}
\subfigure[ ]{\includegraphics[width=3cm]{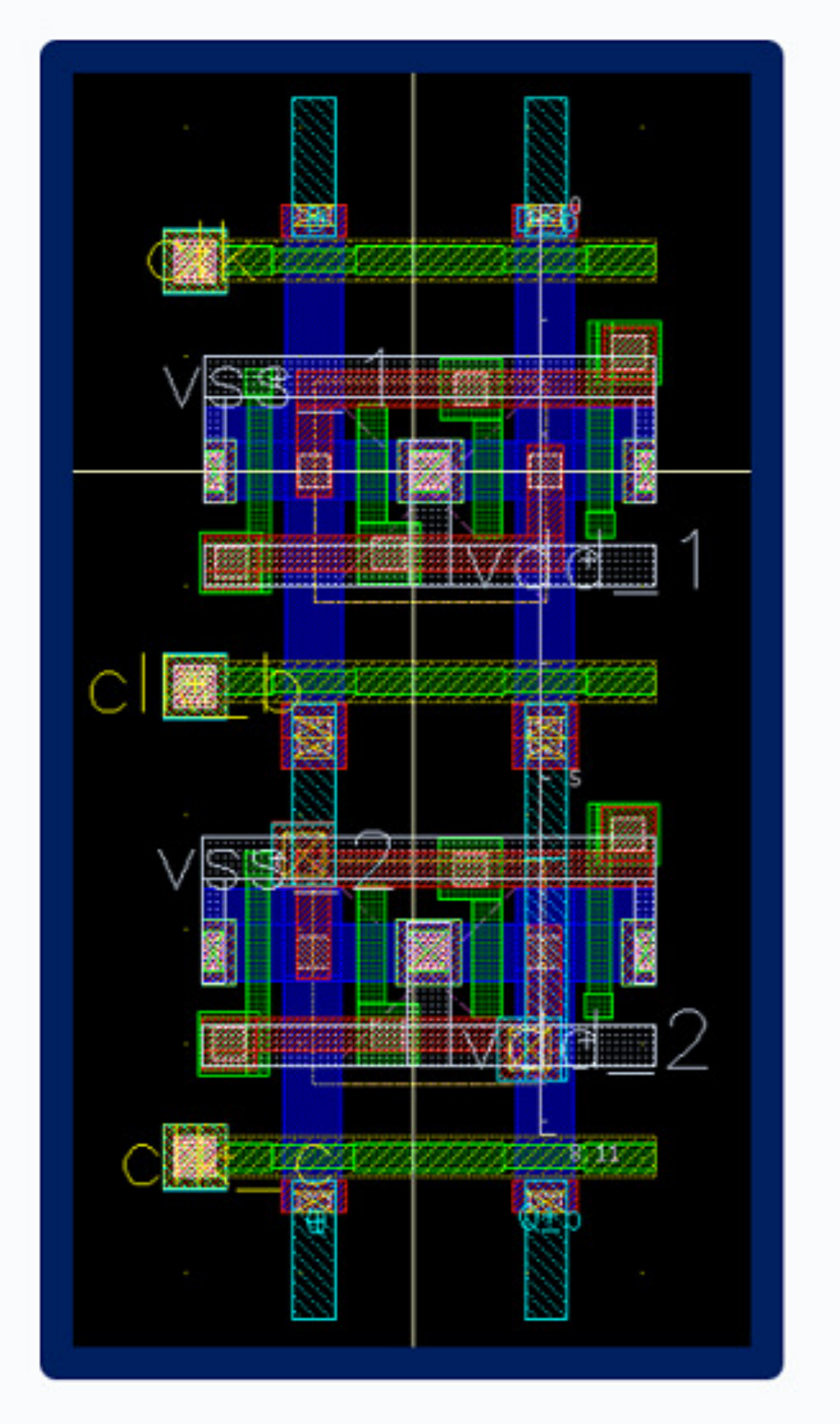}
 \label{fig:fig01right}}
\caption{SRAM cells with dynamic operation: (a) structure of single-port SRAM cell, (b) quasi double-port SRAM cell with dynamic operation, (c) D-FF with dynamic latches. }
\end{figure}

\newpage
Using power-line control by power switches, we cut the drive power of the SRAM cells during writing mode and simultaneously enabled data writing to the multiple memory cells. Our transient analysis of the writing operation using the proposed circuits (Fig. 12) demonstrates that we can achieve simultaneous 32-word SRAM writing with dynamic operation. 

\begin{figure}[htb]
\centering
\includegraphics[height=3in]{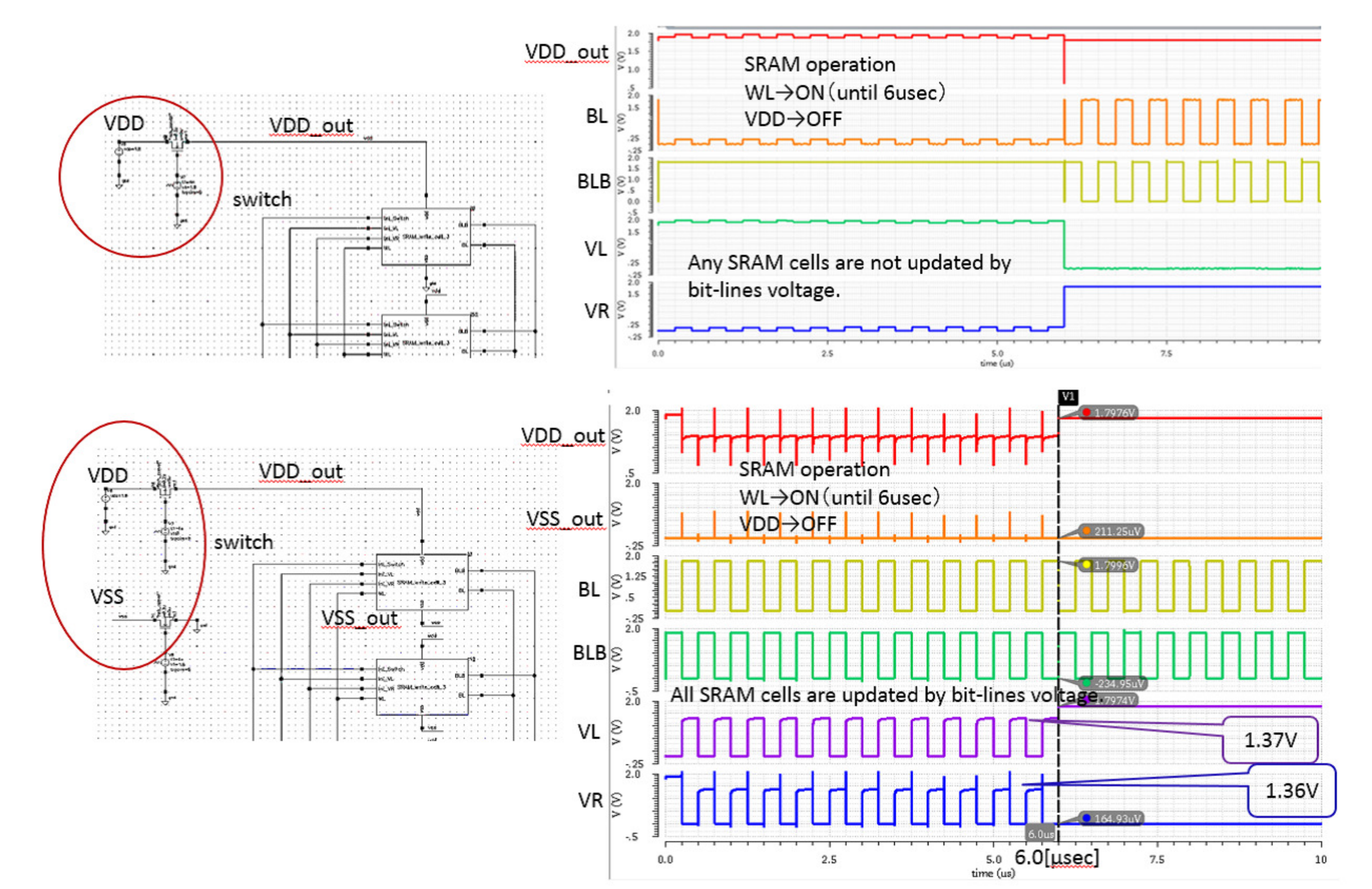}
\caption{Simulation results of simultaneous 32-word SRAM writing. }
\label{fig:magnet}
\end{figure}

Figure 13 shows the layout of a 32$\times$32 pixel array and the pixel circuit with the proposed synchronized TMC, where the ion detector was placed to top of the pixel and the memory cells occupied almost the entire pixel area.
\begin{figure}[htb]
\centering
\includegraphics[height=2.4in]{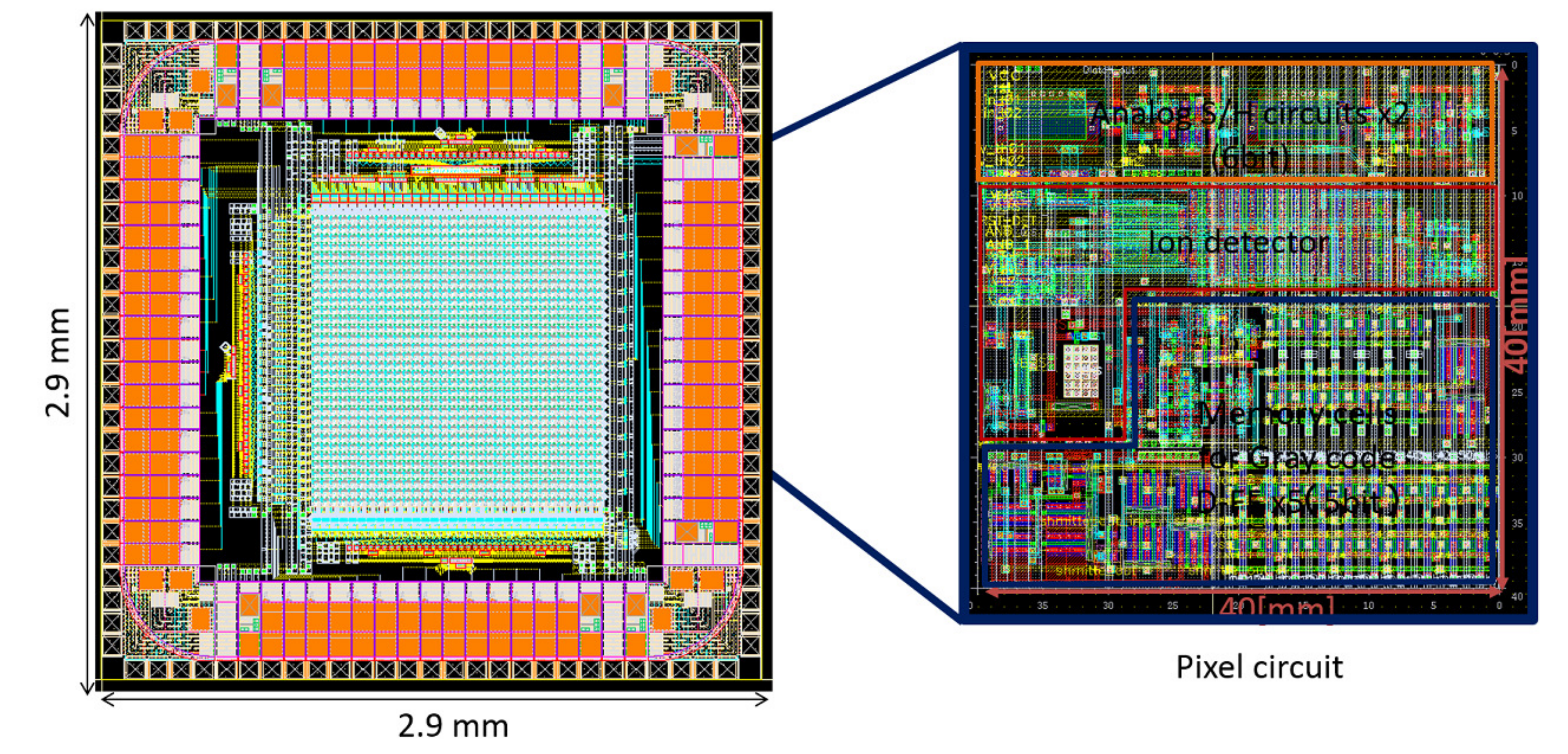}
\caption{Layout of pixel array.}
\label{fig:magnet}
\end{figure}

\clearpage
\section{Conclusion}

We proposed a synchronized TMC/counter architecture to enable time stamping for ion-ToF stigmatic imaging mass spectrometry at a lower clock rate. Using the TMC output for storing the counter state with the Schmitt trigger, we were able to achieve exact causality between the TMC and the counter with robustness against metastability. Also, by applying dynamic operation to SRAM cells, full support of multi-hit/event scenarios can be achieved. The results of transient analysis demonstrate the fully correct synchronous operation at a 100-MHz clock frequency and simultaneous 32-word SRAM writing.

\Acknowledgements
The authors acknowledge the valuable advice and great work by the
personnel of LAPIS Semiconductor Co., Ltd. This work is supported by
JSPS KAKENHI (2510900
315). This work is also supported by VLSI Design
and Education Center (VDEC), the University of Tokyo in collaboration
with Synopsys, Inc., Cadence Design Systems, Inc., and Mentor
Graphics, Inc.









\bibliographystyle{h-physrev5}
\bibliography{watanabe_sjis}

\begin{thebibliography}{1}

\bibitem{hazama_development_2011}
H.~Hazama {\em et~al.},
\newblock Journal of Biomedical Optics {\bf 16}, 046007 (2011).

\bibitem{aoki_novel_2011}
J.~Aoki, H.~Hazama, and M.~Toyoda,
\newblock Journal of the Mass Spectrometry Society of Japan {\bf 59}, 57
  (2011).

\bibitem{arai_time_1998}
Y.~Arai, M.~Ikeno, M.~Sagara, and T.~Emura,
\newblock IEEE Transactions on Nuclear Science {\bf 45}, 735 (1998).

\bibitem{takahashi_digital_2011}
T.~Takahashi {\em et~al.},
\newblock A digital {CDS} scheme on fully column-inline {TDC} architecture for
  an {APS}-{C} format {CMOS} image sensor,
\newblock pp. 90--91, IEEE, 2011.

\bibitem{uchida_12-bit_2014}
D.~Uchida, M.~Ikebe, J.~Motohisa, and E.~Sano,
\newblock A 12-bit, 5.5-{{$\mu$}W} single-slope {ADC} using intermittent
  working {TDC} with multi-phase clock signals,
\newblock in {\em 2014 21st {IEEE} {International} {Conference} on
  {Electronics}, {Circuits} and {Systems} ({ICECS})}, pp. 770--773, 2014.

\end{thebibliography}

\end{document}